\def\year{2018}\relax
\documentclass[letterpaper]{article} %DO NOT CHANGE THIS
\usepackage{aaai18}  %Required
\usepackage{times}  %Required
\usepackage{helvet}  %Required
\usepackage{courier}  %Required
\usepackage{url}  %Required
\usepackage{graphicx}  %Required
\usepackage{verbatim}
\usepackage{multirow}
\usepackage{verbatim}
\usepackage{bm}
\usepackage{color}
\usepackage{amsmath}
\usepackage{amssymb,sansmath}
\usepackage{balance}
\usepackage{longtable}
\usepackage{epstopdf}
\usepackage{epsfig}
\usepackage{subfigure}
\usepackage{xcolor}

\usepackage[ruled]{algorithm2e}
\newcommand{\hide}[1]{} %hide

\newcommand{\vpara}[1]{\vspace{0.1in}\noindent\textbf{#1 }}
\newcommand{\tabincell}[2]{\begin{tabular}{@{}#1@{}}#2\end{tabular}}

\begin{document}

\title{Reinforcement Learning for Relation Classification from Noisy Data}
\author{Jun Feng$^\S$, Minlie Huang$^\S$\thanks{Corresponding author: Minlie Huang, aihuang@tsinghua.edu .cn}, Li Zhao$^\ddag$, Yang Yang$^\dag$, and Xiaoyan Zhu$^\S$ \\
  $^\S$ State Key Lab. of Intelligent Technology and Systems, National Lab. for Information Science and Technology\\
  Dept. of Computer Science and Technology, Tsinghua University, Beijing 100084, PR China \\
  $^\ddag$ Microsoft Research Asia \\
  $^\dag$ College of Computer Science and Technology, Zhejiang University\\
 % $^*$Correspondence Author: aihuang@tsinghua.edu.cn (Minlie Huang)\\
  {\tt feng-j13@mails.tsinghua.edu.cn} ,\quad {\tt aihuang@tsinghua.edu.cn}, \quad{\tt lizo@microsoft.com} \\ 
  {\tt yangya@zju.edu.cn} ,\quad {\tt zxy-dcs@tsinghua.edu.cn}}

\maketitle

\begin{abstract}
Existing relation classification methods that rely on distant supervision assume that a bag of sentences mentioning an entity pair are all describing a relation for the entity pair. Such methods, performing classification at the bag level, cannot identify the mapping between a relation and a sentence, and largely suffers from the noisy labeling problem. 
%%%
In this paper, we propose a novel model for relation classification at the sentence level from noisy data. The model has two modules: an instance selector and a relation classifier. The instance selector chooses high-quality sentences with reinforcement learning and feeds the selected sentences into the relation classifier, and the relation classifier makes sentence-level prediction and provides rewards to the instance selector. The two modules are trained jointly to optimize the instance selection and relation classification processes.
Experiment results show that our model can deal with the noise of data effectively and obtains better performance for relation classification at the sentence level.
\end{abstract}

\section{Introduction}
Relation classification, aiming to categorize semantic relations between two entities given a plain text, is an important problem in natural language processing, particularly for knowledge graph completion and question answering.
Most existing works for relation classification adopt supervised learning approaches, either based on traditional handcrafted features~\cite{mooney2005subsequence,guodong2005exploring} or based on the features automatically generated by deep neural networks  ~\cite{zeng2014relation,DBLP:conf/acl/SantosXZ15}, but all require high-quality annotated data. 

\begin{figure}[t]
 \centering
   \includegraphics[width=0.45\textwidth]{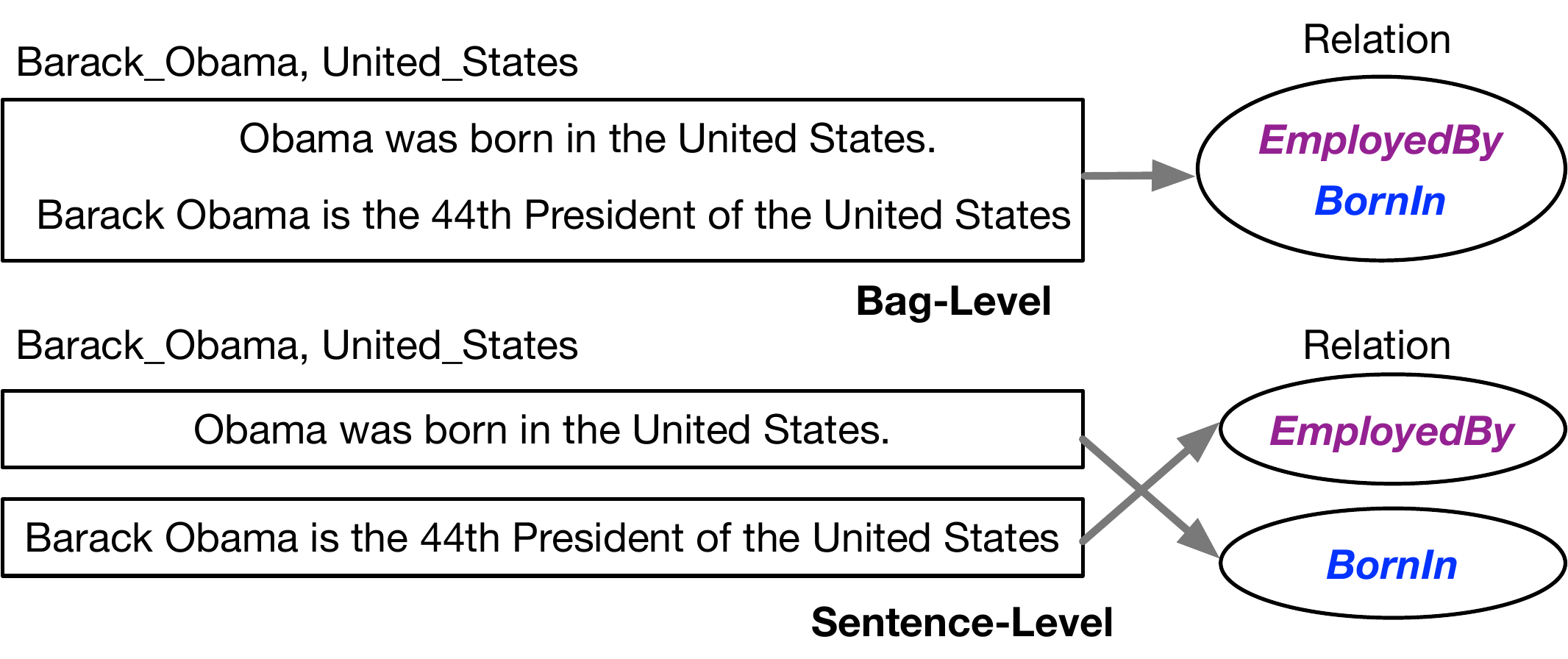}
   %\vspace{-0.1in}
   \caption{Bag-level: Relations are mapped to a bag of sentences, each of which contains the same entity pair; Sentence-level: Each sentence is mapped to a specific relation.}
   %\caption{ a) Bag-level: a sentence bag for two entities contains multiple relations; b)Sentence-level: each sentence is mapped to a specific relation.}
   \label{fig:coarse_grained}
 %\vspace{-0.1in}	
 \end{figure}

%1.2 distant supervision for more data, but also introduce noise
In order to obtain large-scale training data, distant supervision \cite{mintz2009distant} was proposed by assuming that if two entities have a relation in a given knowledge base, all sentences that contain the two entities will mention that relation. 
Although distant supervision is effective to label data automatically, it suffers from the {\it noisy labeling problem}. 
Taking the triple (Barack\_Obama, \textit{BornIn}, United\_States) as an example, the noisy sentence {\it ``Barack Obamba is the 44\textit{th} president of the United State''} will be regarded as a positive instance by distant supervision and a \textit{BornIn} relation is assigned to this sentence, although the sentence does not describe the relation \textit{BornIn} at all. 

To address the issue of noisy labeling, previous studies adopt multi-instance learning to consider the noises of instances~\cite{riedel2010modeling,hoffmann2011knowledge,surdeanu2012multi,zeng2015distant,lin2016neural,DBLP:conf/aaai/Ji0H017}. 
In these studies, the training and test process is proceeded at the bag level, where a bag contains noisy sentences mentioning the same entity pair but possibly not describing the same relation.
%However, in a bag, there may exist noisy sentences that contains the given entities, but does not describe the target relation. 
%In other words, prediction is made for each bag instead of for each sentence.
%Existing works are based on models which jointly select data and extract relation at bag level. 
%Thus model is trained to predict one relation for one sentence bag which may contains noisy sentences. 
%As a result, most of the existing models suffer from two limitations: 1) prediction at the bag level rather than at the sentence level,%, where a bag contains noisy sentences mentioned the same entity pair w.r.t. a relation, 
%2)  incapability of excluding a bag in which all sentences are wrong w.r.t a relation, since the model is trained to predict one relation for one bag without filtering any noisy bag.
As a result, previous studies suffer from two limitations: 
1) Unable to handle the sentence-level prediction; 2) Sensitive to the bags 
%with noisy sentences, especially to those 
with all noisy sentences which do not describe a relation at all. 
%in which all sentences are incorrect. 

%To further explain the limitation of bag-level prediction, Figure~\ref{fig:coarse_grained} gives an example.
%Given two sentences, sentence-level prediction, unlike bag-level prediction, is able to map a particular relation to the corresponding sentence. 
To better explain the first limitation, we show an example in Figure~\ref{fig:coarse_grained}. 
Bag-level prediction can find the two relations ``{\em EmployedBy}'' and ``{\em BornIn}'' between the entity pair ``{\em Barack\_Obama}'' and ``{\em United\_States}''. 
However, sentence-level prediction is able to further map each relation to the corresponding sentences.
As for the second limitation, for each bag, previous bag-level methods retain at least one sentence, even if all the sentences in a given bag are noisy (not describing the relation). Such bags, produced by distant supervision, are quite common.
%in the datasets. 
For instance, our investigation on a widely used dataset\footnote{http://iesl.cs.umass.edu/riedel/ecml/} shows that $53\%$ out of 100 sample bags have no sentences that describe the relation. 
Such noisy bags will definitely decrease the performance of relation classification.

In this paper, to handle the above two limitations, we propose a novel relation classification model consisting of two modules: instance selector and relation classifier. 
%At training time, we train our data selector and relation extractor jointly on sentence bag with noise. We want to train our model in such way, to maximize the average log-likelihood on selected data. 
%The data selector and the relation extractor are updated iteratively.Given data selector, the relation extractor updates its parameter to maximize the average likelihood on selected data, where the likelihood is modeled at sentence level; given relation extractor, the data selector updates its parameter to select data with maximum average likelihood.
By having an explicit instance selector\footnote{Instance is referred to a sentence in this paper.}, we are able to first select high-quality sentences from a sentence bag, and then predict a relation at the sentence level by the relation classifier.
%Thus, our model naturally solve the first limitation. 
To handle the second limitation, our instance selector will filter the entire bag if all sentences are labeled incorrectly.
%For the first limitation, our model can predict the relation with only data selector module at sentence level.
%For the second limitation, our data selector can select no sentences at all from a bag, if the average likelihood of any selected sentence is below certain threshold.
The major challenge here is how to train the two modules jointly, particularly when the instance selector has no explicit knowledge about which sentences are labeled incorrectly.

We address this challenge by casting the instance selection task as a reinforcement learning problem~\cite{sutton1998reinforcement}. 
Intuitively, although we do not have an explicit supervision for the instance selector, we can measure the utility of the selected sentences as a whole. Thus, the instance selection process has the following two properties: first, \textit{trial-and-error-search}, meaning that the instance selector attempts to choose some sentences and obtain feedback (or \textit{reward}) on the quality of the selected sentences from the relation classifier; second, the feedback from the relation classifier can be obtained only when we finish the instance selection process, which is typically \textit{delayed}. These two properties naturally inspire us to utilize 
%to cast the instance selection task as a 
reinforcement learning techniques.
%problem~\cite{sutton1998reinforcement}. 

Our contributions in this work include:
\begin{itemize}
    \item We propose a new model for relation classification, which consists of an instance selector and a relation classifier. This formalization enables our model to extract relations at the sentence level on the cleansed data.
    
    \item We formulate instance selection as a reinforcement learning problem, which enables the model to perform instance selection without explicit sentence-level annotations but just with a weak supervision signal from the relation classifier. 
    
%    \item Experiments show that our model can conduct relation extraction from noisy data at the sentence level with better performance. Furthermore, noisy data can be excluded effectively from a sentence bag.
    %%针对此设计有针对性的实验
\end{itemize}

\section{Related Work}
Relation classification is a common task in natural language processing. Many approaches have been developed, particularly with supervised methods~\cite{mooney2005subsequence,guodong2005exploring,zelenko2003kernel}. However, such supervised methods heavily rely on high-quality labeled data. 

Recently, neural models have been widely applied to relation classification ~\cite{zeng2014relation,DBLP:conf/acl/SantosXZ15,mooney2005subsequence,DBLP:conf/emnlp/YangTMD16} including   convolutional neural networks, recursive neural network~\cite{DBLP:conf/naacl/EbrahimiD15,DBLP:conf/acl/LiuWLJZW15}, and long short-term memory network~\cite{DBLP:conf/acl/MiwaB16,DBLP:conf/emnlp/XuMLCPJ15,DBLP:conf/acl/MiwaB16}. In ~\cite{DBLP:conf/acl/WangCML16}, two levels of attention is proposed in order to better discern patterns in heterogeneous contexts for relation classification.

In general, a large amount of labeled data are required to train neural models, which is quite expensive. To address this issue, distant supervision was proposed ~\cite{mintz2009distant} by assuming that all sentences that mention two entities of a fact triple describe the relation in the triple. In spite of the success of distance supervision, such methods suffer from the noisy labeling issue. To alleviate this issue, many studies formulated relation classification as a multi-instance learning problem~\cite{riedel2010modeling,hoffmann2011knowledge,surdeanu2012multi,zeng2015distant}. In ~\cite{lin2016neural,DBLP:conf/aaai/Ji0H017,liu2017soft}, a sentence-level attention mechanism over multiple instances was proposed and incorrect sentences can be down-weighted. 
However, such multi-instance learning models all predict relations at the bag level but not at the sentence level,
and they can not deal with the bags in which all sentences are not describing a relation at all. 
There are other approaches to reduce the noise of distant supervision using active learning~\cite{sterckx2014using} and negative patterns~\cite{takamatsu2012reducing}.

%There are also direct ways to reduce the noise of distant supervision.
%In ~\cite{sterckx2014using} active learning is proposed to ask for additional human labeling to reduce noisy data. In ~\cite{takamatsu2012reducing} wrong sentences can be removed by negative patterns.%, which are extracted from sentences and are predicted not to express the relations. %However, the patterns are defined by some NLP tools which suffer from error propagation problem.

Previous methods are all at the bag level but not at the sentence level and as such, they cannot find the exact mapping between a relation and a sentence.
Furthermore, these methods are unable to handle the bags in which all the sentences are not describing the relation.
To address these issues, we propose a new framework which first selects correct sentences in the framework of reinforcement learning~\cite{sutton1998reinforcement,narasimhan2016improving} and then predicts relations from each sentence in the cleansed data. 

\section{Methodology}
We propose a new relation classification framework, which is able to select correct sentences from noisy data for better relation classification.
%that are automatically generated by distant supervision.
The proposed framework can predict relations at the sentence level from the cleansed data, rather than at the bag level.
Sentence-level prediction is more friendly to the tasks that need to comprehend sentences such as question answering and semantic parsing. 

Our framework consists of two key modules: the instance selector which selects correct sentences from noisy data, and the relation classifier which predicts relation and updates its parameters with cleaned data. 
The two modules interacts with each other during the training process. 

\subsection{Problem Definition}
%The problem we deal with in this paper is formulated as follows: Given a post $\bm{X} = (x_1,x_2,\cdots, x_{n})$, and an pre-specified emotion category $e\in$ \{{\it Anger, Disgust, Happy, Like, Sad and Other}\} of the generated response, the goal is to generate a response $\bm{Y} = (y_1,y_2,\cdots, y_{m})$ that is coherent to the emotion category $e$. Essentially, the model estimates the probability: $P(\bm{Y}|\bm{X},e)=\prod_{t=1}^{m}P(y_t|y_{<t},\bm{X},e)$.
Formally, we decompose the task of relation classification into two sub-problems in this paper: instance selection and relation classification.

We formulate the instance selection problem as follows: given a set of $<$sentence, relation label$>$ pairs as $X = \{(x_1, r_1), (x_2, r_2), \dots, (x_n, r_n)\}$, where $x_i$ is a sentence associated with two entities $(h_i, t_i)$ and $r_i$ is a noisy relation label produced by distant supervision. The goal is to determine which sentence truly describes the relation and should be selected as a training instance.
%the \textit{action} $a_i \in \{0, 1\} $, where $a_i = 1$ denotes that $(x_i, r_i)$ 
%is a positive instance and 
%will be selected as a training instance, at the current \textit{state} $s_i$, where $s_i$ is a representation vector corresponding to the already chosen instances and the current instance $x_i$. 
%Essentially, the model estimates the probability: $P_\Theta(a|s)$.

The relation classification problem is formulated as follows: given a sentence $x_i$ and the mentioned entity pair $(h_i, t_i)$, the goal is to predict the semantic relation $r_i$ in $x_i$. 
Essentially, the model estimates the probability: $p_\Phi(r_i | x_i, h_i, t_i)$.
%We put the instances with the same corresponding entity pair together and call is bag. Thus, we reorganize the data as below. There are $N$ bags $B = \{B_1, B_2, \dots, B_N\}$ and that the \textit{i}-th bag contains $|B_i|$ sentences $B_i = \{b_i^1, b_i^2, \dots, b_i^{|B_i|} \}$. The sentence in the \textit{i}-th bag are with the same corresponding entity pairs$(h_i, t_i)$ and describe the same relation type $rt_i$.
\subsection{Overview}
\begin{figure}[t]
 \centering
   \includegraphics[width=0.45\textwidth]{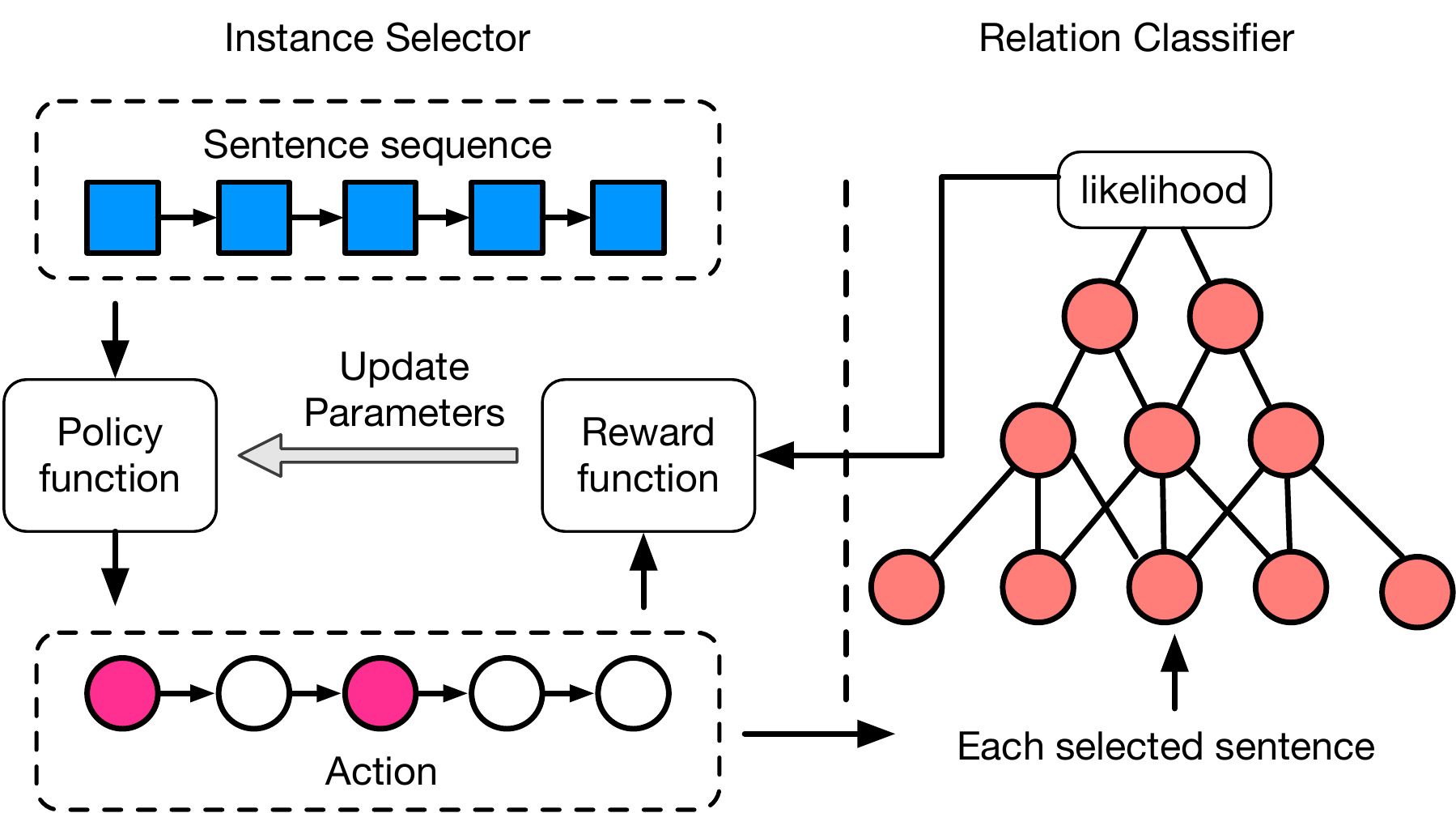}
   %\vspace{-0.1in}
   \caption{Overall process. The instance selector chooses sentences according to a policy function, and then the selected sentences are used to train a better relation classifier. The instance selector updates its parameters, with a reward computed from the relation classifier.}
   \label{fig:framework}
 %\vspace{-0.1in}	
 \end{figure}

%The data selector and the relation extractor are updated iteratively.Given data selector, the relation extractor updates its parameter to maximize the average likelihood on selected data, where the likelihood is modeled at sentence level; given relation extractor, the data selector updates its parameter to select data with maximum average likelihood.
The proposed model is based on a reinforcement learning framework and consists of two components: the \textit{instance selector} and the \textit{relation classifier}. 
In the instance selector, each sentence $x_i$ has a corresponding \textit{action} $a_i$ to indicate whether or not $x_i$ will be selected as a training instance for relation classification. 
The state $s_i$ is represented by the current sentence $x_i$, the already chosen sentences among $\{x_1, \cdots, x_{i-1}\}$, and the entity pair $h_i$ and $t_i$ in sentence $x_i$.
The instance selector samples an action given the current state according to a stochastic policy. 
For the relation classifier, it adopts a convolutional architecture to automatically determine the semantic relation for an entity pair in a given sentence. The
instance selector distills the training data to the relation classifier to train the convolutional neural network. 
Meanwhile, the relation classifier gives feedback to the instance selector to refine its policy function. 
Figure~\ref{fig:framework} gives an illustration of how the proposed framework works.

With the help of the instance selector, our method directly filters out noisy sentences. Unlike reducing the weights of noisy sentences~\cite{lin2016neural} or retaining one sentence in a bag~\cite{zeng2014relation}, our method is better at dealing with noisy data. 
The relation classifier is trained and tested at the sentence level on the cleansed data, whereas previous models treat the sentence bag as a whole and predict relation at the bag level.
%% 这个地方还需要说一下我们的方法和其它人的方法有什么不同，有什么相同，给人更深的印象

\subsection{Instance Selector}
We cast instance selection as a reinforcement learning problem. 
The instance selector is the agent, who interacts with the environment that consists of data and the relation classifier. 
The agent follows a policy to decide which action (choosing the current sentence or not) at each state (consisting of the current sentence, the chosen sentence set, and the entity pair), and then receive a reward from the relation classifier at the terminal state when all the selections are made. 

As aforementioned, we can obtain a delayed reward from the relation classifier only when the selection on all the training instances are finished. Thus, we can only update the policy function once for each scan of the entire training data, which is obviously inefficient.
To obtain more feedbacks and to make the training process more efficiently, we split the training sentence instances $X = \{x_1, \dots, x_n\}$ into $N$ bags $\mathbf{B} = \{B^1, B^2, \dots, B^N\}$ and compute a reward when we finish data selection in a bag. Each bag corresponds to a distinct entity pair, and each bag $B^k$ is a sequence of sentences $\{ x_1^k, x_2^k, \dots, x_{|B^k|}^k \}$ with the same relation label $r^k$, however, the relation label is noisy. 
%Formally, we formulate the instance selection problem as a Markov decision process (MDP). 
%Therefore, we put the instances into $N$ bags $B = \{B_1, B_2, \dots, B_N\}$, where each bag corresponds to a distinct entity pair, and each bag $B_i$ is a sequence of instances $\{ x_i^1, x_i^2, \dots, x_i^{|B_i|} \}$. 
We define the action as selecting a sentence or not according to a policy function. 
The reward is computed once the selection decisions are completed on one bag. %Table~\ref{fig:rl} shows an example of the sequential decision problem with one bag.
%In detail, we represent the MDP as a tuple $<s, a, T, r>$, where $s$ is the state, $a$ represents the action and $r = r(s, a)$ is the reward function, and $T(s'|s, a)$ is the transition function.
%%%
%% hml：这个地方，selector是以bag为单位的，但是rel classifier是以句子为单位的，通篇都不是很清楚如何从bag --》 sentence 这个切换；要在overview 或者 某个地方说得更加清晰！！！
When the training process of the instance selector is completed, we merge all the selected sentences in each bag to obtain a cleansed dataset $\hat{X}$.
Then, the cleansed data will be used to train the relation classifier at the sentence level.

We will introduce (i.e., \textit{state}, \textit{action}, and \textit{reward}) as follows. To be clear, we will omit the superscript $k$ which denotes the bag index. Thus, the formulation hereafter is based on only one bag.

%%请把所有的向量加粗！！

%\subsubsection{State}
\vpara{State.}
The state $s_i$ represents the current sentence, the already selected sentences, and the entity pair when making decision on the $i$-th sentence of the bag $B$. We represent the state as a continuous real-valued vector
%% 这个F实在令人不爽
$\bm{F}(s_i)$, which encodes the following information:
%%前后顺序要对应！！
1) The vector representation of the current sentence, which is obtained from the non-linear layer of the CNN for relation classification;
2) The representation of the chosen sentence set, which are the average of the vector representations of all chosen sentences;
3) The vector representations of the two entities in a sentence, obtained from a pre-trained knowledge graph embedding table.

\vpara{Action.}
We define an action $a_i \in \{0, 1\}$ to indicate whether the instance selector will select the $i$-th sentence of the bag $B$ or not.
%For instance selection, the action space includes two actions $a \in \{0, 1\}$, denoting whether the instance selector selects the current instance or not. 
We sample the value of $a_i$ by its policy function $\pi_\Theta(s_i, a_i)$, where $\Theta$ is the parameters to be learned. In this work, we adopt a logistic function as the policy function:
\begin{equation}
\begin{aligned}
\pi_\Theta(s_i, a_i) & = P_\Theta(a_i|s_i) \\ & = a_i \sigma(\bm{W} *\bm{F}(s_i) + \bm{b}) \\ & + (1 - a_i) (1 - \sigma(\bm{W} *\bm{F}(s_i) + \bm{b}))
\end{aligned}
\end{equation}
\noindent where $\bm{F}(s_i)$ is the state feature vector, and $\sigma(.)$ is the sigmoid function with the parameter $\Theta = \{\bm{W}, \bm{b}\}$.

\vpara{Reward.}
The reward function is an indicator of the utility of the chosen sentences. 
For certain bag $B =\{x_1, \dots, x_{|B|}\}$, we sample an action for each sentence, to determine whether the current sentence should be selected or not. We assume that the model has a terminal reward when it finishes all the selection. Therefore we only receive a delayed reward at the terminal state $s_{|B| + 1}$. The reward is zero at other states. Therefore,  the reward is defined as follows:
\begin{equation}
r(s_i|B) = 
\begin{cases}
0& i < |B| + 1 \\
\frac{1}{|\hat{B}|}\sum\limits_{x_j \in \hat{B}} \log p(r | x_j)& i = |B| + 1 
\end{cases}
\end{equation}
where $\hat{B}$ is the set of selected sentences, which is a subset of $B$, and $r$ is the relation label of bag $B$. As shown in Figure  \ref{fig:framework}, $p(r | x_j)$ is calculated by the relation classifier which is given by a CNN model. For the special case $\hat{B}=\emptyset$, we set the reward as the average likelihood of all sentences in the training data, which enables our instance selector to exclude noisy bag effectively.

Note that the relation classifier is at the sentence-level since it computes $p(r|x)$ for each sentence. The reward is computed on a new bag of sentences selected by the instance selector.
Essentially, the above reward evaluates the overall utility of all the actions made by the policy. It supervises the instance selector to maximize the average likelihood of the chosen instances, which makes the objective function of the instance selector consistent with the relation classifier.

In the selection process, not only the final action contributes to this reward, but also all the previous actions do. Therefore, this reward is delayed, and can be handled very well by reinforcement learning techniques \cite{sutton1998reinforcement}.
%Intuitively, the reward definition is designed to how evaluate how easy can the data be classified by our CNN relation extractor, namely, data with lower average loss. This design is partially inspired by attention mechanism, where we select data aiming at minimize the overall loss.
%\textit{===revised by Li Zhao ends here===}

\vpara{Optimization.}
% this objective should be for all bags, rather than particular one.
%\textit{===revised by Li Zhao starts here===}
For a bag $B$, we aim to maximize the expected total reward. More formally, our objective function is defined as
\begin{equation}
\begin{aligned}
    J(\Theta) & = V_{\Theta}(s_1|B) \\ &= E_{s_1, a_1, s_2, \dots, s_i, a_i, s_{i+1}\dots} [\sum_{i = 0}^{|B|+1} r(s_i|B)]
\end{aligned}
\end{equation}
where $a_i \sim \pi_\Theta(s_i, a_i)$, $s_{i+1} \sim P(s_{i+1}|s_{i}, a_i)$. The transition functions $P(s_{i+1}|s_{i}, a_i)$ are equal to 1, since the state $s_{i+1}$ is fully determined by the state $s_i$ and $a_i$.  $V_{\Theta}$ is the value function, and $V_{\Theta}(s_1|B)$ represents the expected future total reward that we can obtain by starting at certain state $s_1$ following policy $\pi_{\Theta}(s_i,a_i)$.
%
%where $V(s_0^k)$ is the total expected reward by following policy $\pi_{\Theta}(s_i^k,a_i^k)$ for the $k$-th bag. %$V(s_0^k)$ is the value function and is the expected future total reward that we can obtain by starting at state $s_0^k$ for bag $B^k$ and following policy $\pi_{\Theta}(s_i^k,a_i^k)$. 
%Formally, since we only have a delayed reward, which is a non-zero reward at the terminal state, the total expected rewards at each selection state are all the same. Therefore, we define this same value $V_{\Theta}(s_i)$ for bag $B_i$ as follows:
%

According to the policy gradient theorem~\cite{PolicyGradient} and the REINFORCE algorithm~\cite{williams1992simple}, we compute the gradient in the following way.
For each bag $B$, we sample an action for each state sequentially according to the current policy.
We then get a sampled trajectory $\{s_1, a_1, s_2, a_2,..., s_{|B|}, a_{|B|}, s_{|B|+1}\}$ and a corresponding terminal reward $r(s_{|B|+1} | B)$. 
Since we only have a non-zero terminal reward, the value function is the same for all states from $s_1$ to $s_{|B|}$, namely $v_i = V(s_i|B) = r(s_{|B|+1} | B)$, for $i = 1,2,..., |B|$. 
We update the current policy using the following gradient:
\begin{equation}
\Theta \leftarrow \Theta + \alpha \sum_{i=1}^{|B|}v_i \nabla_\Theta \log \pi_\Theta(s_i, a_i)
\end{equation}

%%% 我觉得这个地方，要修改为一个instance extractor 和 relation extractor的迭代算法 
\begin{algorithm}[!t]
\begin{enumerate}
\setlength{\itemsep}{0pt}
\setlength{\parsep}{0pt}
\setlength{\parskip}{0pt}
\item Initialize the parameters of the CNN model of relation classifier and the policy network of instance selector with random weights respectively
\item Pre-train the CNN model to predict relation $r_i$ given the sentence $x_i$ by maximizing $\log p(r_i|x_i)$
\item Pre-train the policy network by running Algorithm ~\ref{alg:paper} with the CNN model fixed.
\item Run Algorithm ~\ref{alg:paper} to jointly train the CNN model and the policy network until convergence
\end{enumerate}
\caption{Overall Training Procedure}
\label{alg:complete}
\end{algorithm}
\begin{algorithm}[!t]
\KwIn{Episode number $L$. Training data $\mathbf{B} = \{B^1, B^2, \dots, B^N\}$. A CNN and a policy network model parameterized by $\Phi$ and $\Theta$, respectively} 
Initialize the target networks as: $\Phi' = \Phi, \Theta' = \Theta$ \\
\For {episode $l = 1$ to $L$}
{
	Shuffle $\mathbf{B}$ to obtain the bag sequence $\mathbf{B} = \{B^1, B^2, \dots, B^N\}$ \\
	\ForEach {$B^k \in \mathbf{B}$}
	{
		Sample instance selection actions for each data instance in $B^k$ with $\Theta'$:\\
		(To be clear, we omit the superscript $k$ below)
		\qquad $A = \{a_1, \dots, a_{|B|}\}, a_{i}\sim \pi_{\Theta'}(s_i, a_i)$ \\
		%Compute reward $rw_i$ with $\Phi'$\\
		Compute delayed reward $r(s_{|B| + 1} | B)$ \\
		Update the parameter $\Theta$ of instance selector:
		\quad $\Theta \leftarrow \Theta + \alpha \sum\limits_i v_i \nabla_{\Theta} \log \pi_{\Theta}(s_i, a_i)$, where $v_i = r(s_{|B| + 1} | B)$
	}
	Update $\Phi$ in the CNN model\\
	Update the weights of the target networks:\\
	    \quad $\Theta' = \tau \Theta + (1 - \tau) \Theta'$ \\
	    \quad $\Phi' = \tau \Phi + (1 - \tau) \Phi'$
}
\caption{Reinforcement Learning Algorithm for the Instance Selector}
\label{alg:paper}
\end{algorithm}
\subsection{Relation classifier}
In the relation classifier, we adopt a CNN architecture to predict relations.
The CNN network has an input layer, a convolution layer, a max pooling layer and a non-linear layer from which the representation is used for relation classification.
%In relation extraction, we transform the instance into a distributed representation by a CNN. 
%Four parts are included in this procedure: input representation, convolution, max pooling, non-linear transformation and softmax output.

\vpara{Input layer.}
For each sentence $x$, we represent it as a list of vectors $\textbf{x} = (\mathbf{w}_1, \mathbf{w}_2, \dots, \mathbf{w}_m)$. Each representation vector consists of two parts: one is the word embedding; the other is the position embedding. Word embeddings are obtained from word2vec\footnote{https://code.google.com/p/word2vec/}, and the dimension is $d^w$. 
%are encoded  by an embedding matrix $\mathbf{V} \in \mathbb{R}^{d^w \times |V|}$, where $V$ is a fixed-sized vocabulary. 
Similar to ~\cite{zeng2014relation}, we use $d^p$-dimensional position embeddings, which are vector representations of the relative distances from the current word respectively to the head or tail entities in this sentence. %Position embeddings map the relative distances (integers) to a $d^e$-dimension vector.
We concatenate the word and position embeddings of each word to form a new vector $\textbf{w}_i$ ($\mathbf{w}_i \in \mathbb{R}^d$, and $d = d^w + 2 \times d^p$), and then input these vectors to the CNN model.

\vpara{CNN.}
In order to obtain high-level and abstractive representation of the raw input of a sentence, we apply a CNN structure for relation classification. This can be briefly described as below:
%As the relation extraction task is to predict a relation label for an input sentence not for each words, we need to utilize all local features and perform the prediction globally. To merge these features, we use a convolutional method:
%
\begin{equation}
    \mathbf{L} = \text{CNN}(\mathbf{x})
\end{equation}
where $\textbf{x}$ is the input vectors as described in the input layer and $\mathbf{L} \in \mathbb{R}^{d^s}$ is the output of the max pooling layer.
In this structure, there is a convolution layer, and a max pooling layer. The convolution operation is performed on 3 consecutive words, and the number of feature maps $d^s$ is set to $230$, the same as the setting of \cite{lin2016neural}. Hence, the convolution parameters are $\mathbf{W}_{f} \in \mathbb{R}^{d^s \times (3d)}$ and $\mathbf{b}_f \in \mathbb{R}^{d^s}$.

Then, the probability for relation prediction $p(r | x; \mathbf{\Phi})$ is given as follows:
\begin{equation}
p(r|x; \mathbf{\Phi}) = softmax(\mathbf{W}_r*tanh(\bm{L}) + \mathbf{b}_r)
\end{equation}
where $\mathbf{W}_r \in \mathbb{R}^{n_r \times d^s}$ and $\mathbf{b}_r \in \mathbb{R}^{n_r}$ are parameters in the fully-connected layer, $n_r$ is the total number of relation types, and $\mathbf{\Phi}=\{\mathbf{W}_f, \mathbf{b}_f, \mathbf{W}_r,\mathbf{b}_r\}$.

The key difference between our relation classifier and other studies lies in that our classifier performs relation classification at the sentence level.
The input to the relation classifier in other studies is a bag of sentences.
Instead, the input to ours is just one sentence, since we already filter out noisy sentences with the instance selector.

\vpara{Loss function.}
Given the selected training set $\{\hat{X}\}$ 
%$\{(x_i, r_i)\}$ 
provided by the instance selector, we define the objective function of the relation classifier using cross-entropy as follows:
\begin{equation}
\mathcal{J}(\Phi) = -\frac{1}{|\hat{X}|}\sum_{i = 1}^{|\hat{X}|} \log p(r_i|x_i; \Phi)
\label{eq:cnn_loss}
\end{equation}
%where $\Phi$ indicates parameters of the relation extractor.

\subsection{Model Training}
As the instance selector and the relation classifier are correlated mutually, we train them jointly.% to obtain better performance. 
The complete joint training process is described in Algorithm~\ref{alg:complete}. 
To optimize the policy network in the instance selector, we use a Monto-Carlo based policy gradient method~\cite{williams1992simple}, which favors actions with high sampled reward.
To optimize the CNN component, we use a gradient descent method to minimize the objective function (i.e., Eq. \ref{eq:cnn_loss}).
We pre-train the model before the joint training process starts. 
%The reason of the pre-training is that, starting the training with randomly initialized parameters may lead to meaningless training, as neither the instance selector and relation extractor would provide adequate training signals for each other.
We first pre-train the CNN in the relation classifier, and then pre-train the policy function by computing the reward with the pre-trained CNN, while the parameters of the CNN model are frozen. 
At last, we jointly train the instance selector and the relation classifier.
We found such a pre-training strategy is quite crucial for our method, which is also widely recommended by many other reinforcement learning studies\cite{bahdanau2016actor}. 

Algorithm~\ref{alg:paper} presents the details of the joint training process. The relation classifier provides a mechanism of computing the rewards of the selected sentences to refine the instance selector. 
The instance selector chooses high-quality data by excluding wrongly labeled sentences to better train the relation classifier.
In order to have a stable update, we take advantage of a target policy network and a target CNN with parameter sets $\Theta'$ and $\Phi'$ respectively, similar to ~\cite{lillicrap2015continuous}. The parameters in the target networks are updated much more slowly than the original ones. We update $\Theta'$ and $\Phi'$ by linear interpolation: $\Theta' \leftarrow (1 - \tau) \Theta' + \tau \Theta$ and $\Phi' \leftarrow (1 - \tau) \Phi' + \tau \Phi$, where $\tau \ll 1$ is a hyper-parameter. 

\section{Experiment}
\subsection{Experiment Setup}
\vpara{Dataset.}
To evaluate our model, we adopted a widely used dataset\footnote{http://iesl.cs.umass.edu/riedel/ecml/} generated by the sentences in NYT\footnote{New York Times, a widely used text corpus.} and developed by~\cite{riedel2010modeling}.
There are 522,611 sentences, 281,270 entity pairs, and 18,252 relational facts in the training data; and 172,448 sentences, 96,678 entity pairs and 1,950 relational facts in the test data.
Among the data, there are 39,528 unique entities and 53 unique relations from Freebase including a special relation {\em NA} that signifies no relation between two entities in a sentence.

\vpara{Word and entity embedding.}
We adopted word2vec to train the word embeddings on the NYT corpus. For entity embedding, we implemented the TransE model ~\cite{bordes2013translating} and trained it on a set of Freebase
%\footnote{Freebase was a large collaborative knowledge base consisting of data composed mainly by its community members.} 
fact
triples whose entities have been mentioned in the training and test data.

\vpara{Model pre-training.}
As described in Algorithm~\ref{alg:paper}, we pre-trained the relation classifier and instance selector before the joint training process. As the reward is calculated based on the CNN model in the relation classifier, we first pre-trained the CNN model on the entire training data. Then, we fixed the parameters of the CNN model and pre-trained the policy function in the instance selector where the reward is obtained from the fixed CNN model.
%%it's better what the data is for training the 

\vpara{Parameter setting.}
Similar to previous studies, we tuned our model using three-fold cross validation. For the parameters of the instance selector, we set the dimension of entity embedding as $50$, the learning rate as $0.02$/$0.01$ at the pre-training stage and joint training stage respectively.  The delay coefficient $\tau$ is $0.001$. 

For the parameters of the relation classifier, the word embedding dimension $d^w = 50$ and the position embedding dimension $d^p = 5$. The window size of the convolution layer $l$ is $3$. The learning rate of the instance selector is $\alpha = 0.02$ both at the pre-training and joint training stage. The batch size is fixed to $160$. The training episode number $L = 25$. We employed a dropout strategy with a probability of $0.5$ during the training of the CNN component.
%In the implementation of CNN in the relation classifier, we employ a dropout strategy on the output layer to prevent over-fitting, which randomly sets the hidden units to zero with a probability of $0.5$ during training.

\subsection{Sentence-Level Relation Classification}
As discussed previously, the key difference between our method and other models lies in that our method can perform sentence-level relation classification. 
We conducted manual evaluation on relation classification in this section.

\vpara{Evaluation settings.}
We predicted a relation label for each sentence, instead of for each bag. For example, the task in Figure~\ref{fig:coarse_grained} needs to map the first sentence to relation ``{\em BornIn}'' and the second sentence to ``{\em EmployedBy}''.

%As shown in Figure~\ref{fig:coarse_grained}, in this task, it needs to map the first sentence to relation ``{\em BornIn}'' and the second sentence to ``{\em EmployedBy}''.

%
%However, 
Since the data obtained from distant supervision are noisy, we randomly chose 300 sentences and manually labeled the relation type for each sentence to evaluate the classification performance. We adopted accuracy and macro-averaged $F_1$ as the evaluation metric. 

\vpara{Baselines.}
We adopted three state-of-the-art baselines:
\begin{itemize}
    \item \textbf{CNN}~\cite{zeng2014relation} is a sentence-level classification model. It does not consider the noisy labeling problem. %For each sentence with noisy label, it adopts an end-to-end convolutional neural network to extract latent features and predict the relation using the extracted features.
    \item \textbf{CNN+Max}~\cite{zeng2015distant} is a bag-level classification model. It assumes that there is one sentence describing the relation in a bag. It chooses the most correct sentence in each bag. 
    \item \textbf{CNN+ATT}~\cite{lin2016neural} is also a bag-level model, similar to CNN+Max. It adopts a sentence-level attention over the sentences in a bag and thus can down weight noisy sentences in a bag.
\end{itemize}
 
CNN is a sentence-level model that is trained directly on noisy data.
For bag-level models (CNN+Max and CNN+ATT), the training process is the same as the referenced papers. 
During test, each sentence is treated as a bag and a relation is predicted for each bag. In this scenario, the bag-level relation prediction is exactly the same as the sentence-level prediction.
All the baselines were implemented with the source codes released by ~\cite{li2016deep}.

\begin{table}
\small
\centering
\begin{tabular}{| c | c | c|} \hline
Method & Macro $F_1$ & Accuracy \\ \hline
CNN & 0.40 & 0.60\\
CNN+Max & 0.06 & 0.34\\
CNN+ATT & 0.29 & 0.56\\ \hline
CNN+RL(ours) & \textbf{0.42} & \textbf{0.64}\\ \hline
\end{tabular}
\caption{Performance on sentence-level relation classification.}
\label{tb:f1}
\end{table}

\vpara{Results.}
Results in Table~\ref{tb:f1} reveal the following observations. 
% From the results, we see that our model achieves a $64.6\%$ improvement comparing with baselines. 
% In particular, CNN does not consider the noisy labeling problem and thus under-performs our model. 
% Interestingly, CNN outperforms CNN+Max and CNN+ATT, which try to solve the noisy labeling problem.
% It suggests the difficulty of the instance selection problem. 
% However, our method addresses this problem well by formulating it to a reinforcement learning problem.  
%We have these observations: 
\begin{itemize}
    \item CNN+RL obtains superior performance than CNN, indicating that filtering noisy data by instance selection benefits the task. 
    \item CNN+RL outperforms CNN+Max and CNN+ATT remarkably. It shows the effectiveness of instance selection with reinforcement learning. 
    \item The sentence-level models (CNN and CNN+RL) perform much better than the bag-level models (CNN+Max and CNN+ATT), indicating that bag-level models do not perform well for sentence-level prediction.
\end{itemize}

%The results show that our proposed model CNN+RL can extract sentence-level relations on distant supervision data which contains noisy-label instances, whereas the other sentence-level model CNN cannot deal with noisy-label problem and bag-level models CNN+Max and CNN+ATT unable to handle the sentence-level relation prediction well.

\subsection{Instance Selection}
We then evaluated the effectiveness of our instance selector from several aspects. First, we evaluated whether the selected data by our instance selector are better for relation classification. Second, we justified the accuracy of selection decision in the selector by manually checking the decisions on sentences. 
Third, we compared the proposed RL selection strategy in our selector with greedy selection. Last, we assessed whether the selector has the ability of filtering those bags that contain all noisy sentences.

%These extensive experiments can evaluate the effectiveness of our instance selector completely.

%
\begin{figure}[t]
 \centering
   \includegraphics[width=0.33\textwidth]{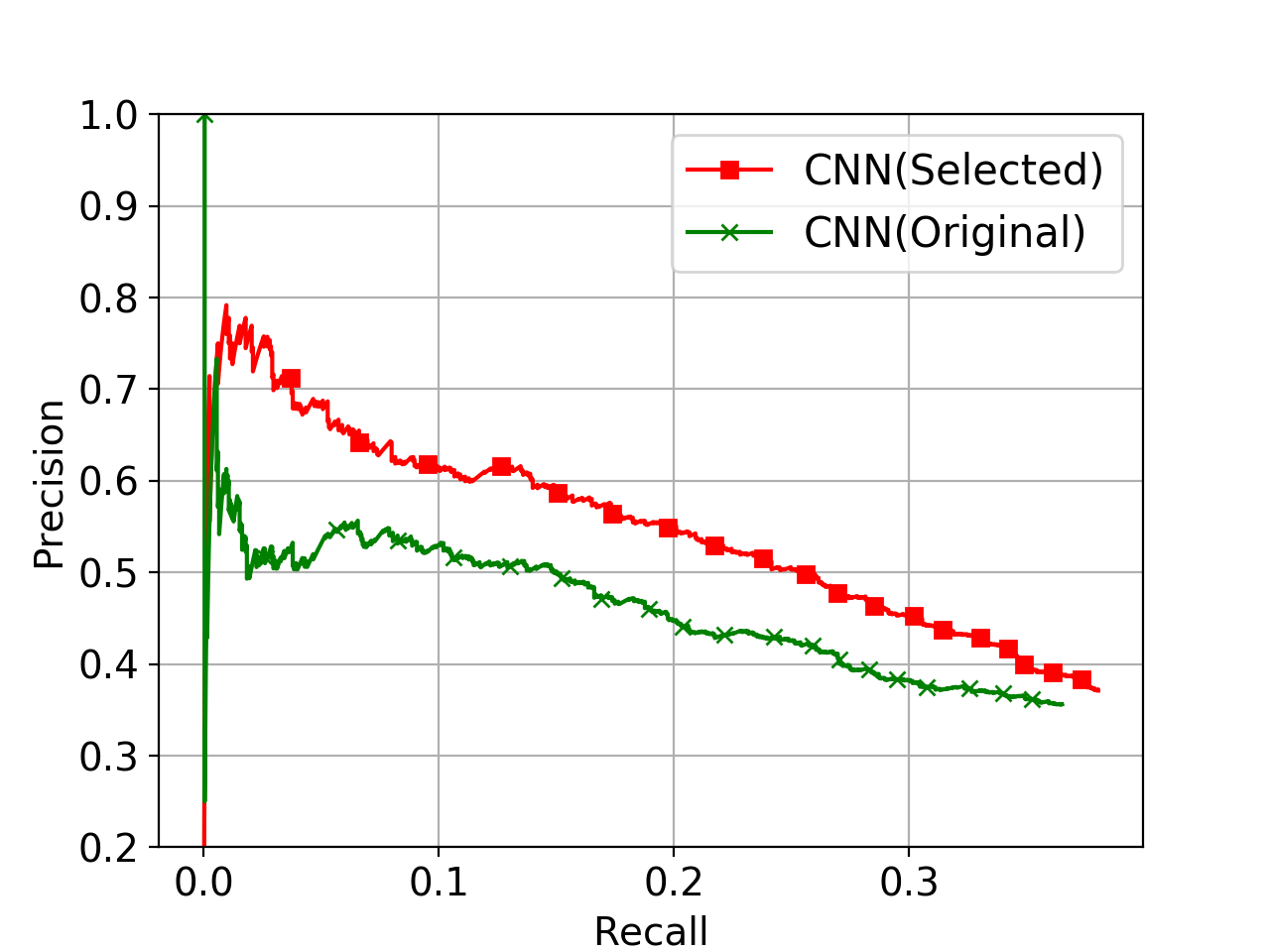}
   %\vspace{-0.1in}
   \caption{Comparison between the CNN model trained on the original and selected data.}
   \label{fig:cnn+rl}
 %\vspace{-0.1in}	
    \includegraphics[width=0.33\textwidth]{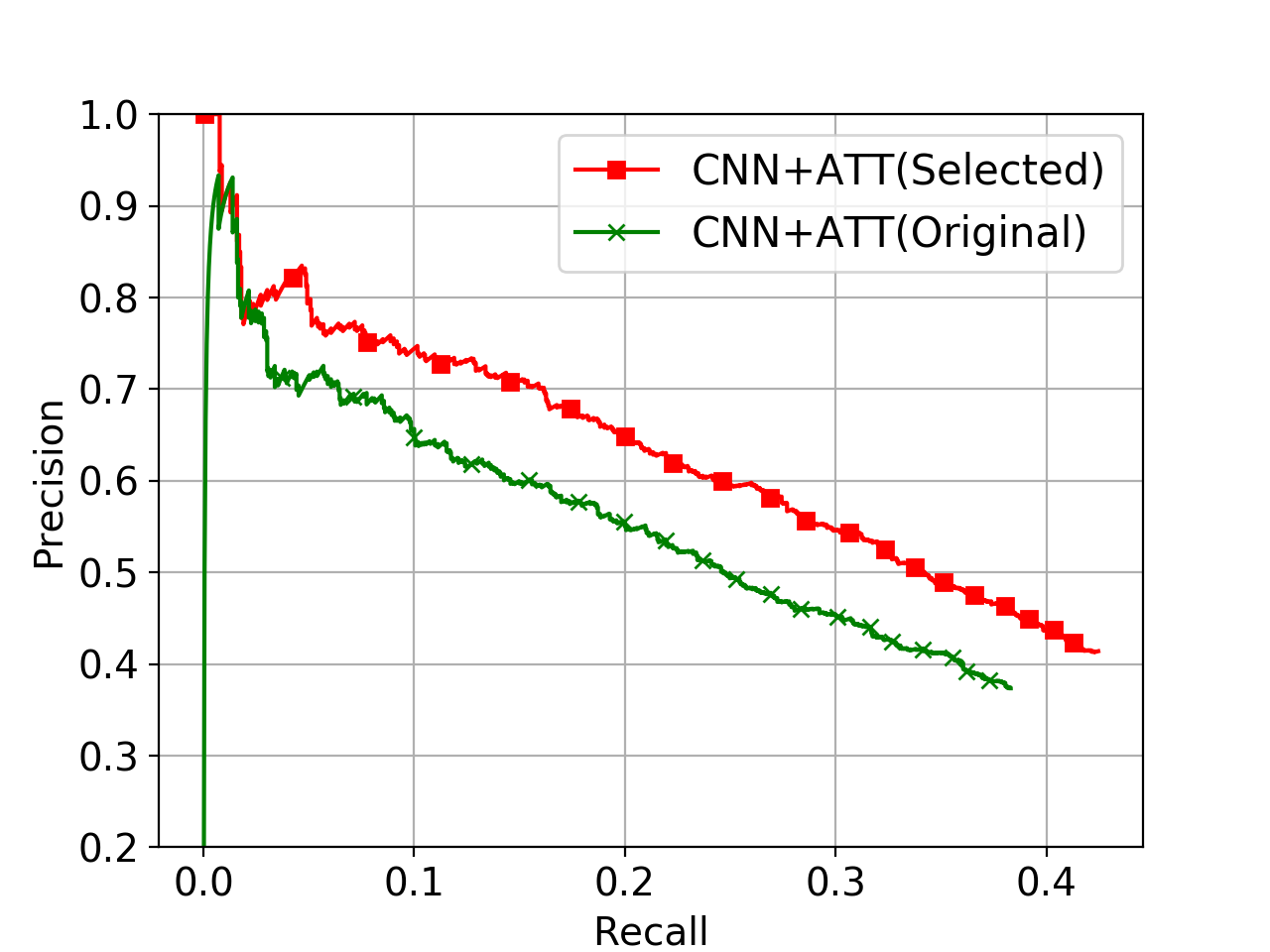}
   %\vspace{-0.1in}
   \caption{Comparison between the CNN+ATT model trained on the original and selected data. }
   \label{fig:cnn+att+rl}
 %\vspace{-0.1in}	
\end{figure}

\begin{table*}
\small
\centering
\begin{tabular}{| l | c | c | c |}\hline
\textbf{Bag I} (\textbf{Entity Pair}: fabrice\_santor, france; \textbf{Relation}:/people/person/nationality) & CNN+RL & CNN+ATT & CNN+Max\\ \hline
\tabincell{l}{though not without some struggle, federer, the world 's top-ranked player, advanced to the fourth \\ round with a thrilling, victory over the crafty \textbf{fabrice\_santoro} of \textbf{france}, who is ranked 76th.}
 & 1 & 0.60 & 0 \\ \hline
in his quarterfinal , nalbandian overwhelmed unseeded \textbf{fabrice\_santoro} of \textbf{france} & 1 & 0.39 & 1 \\ \hline
\tabincell{l}{\textbf{fabrice\_santoro}, 33 , of \textbf{france} finally reached the quarterfinals in a major on his 54th attempt by \\ defeating the 11th-seeded spaniard david ferrer} & 1 & 0.01 & 0 \\ \hline \hline
\textbf{Bag II} (\textbf{Entity Pair}: jonathan\_littel, france; \textbf{Relation}:/people/person/nationality) & & & \\ \hline
\tabincell{l}{\textbf{jonathan\_littell}, a new york-born writer whose french-language novel about a murderous\\ and degenerate officer has been the sensation of the french publishing season, on monday \\ became the first american to win \textbf{france}'s most prestigious literary award, the prix goncourt } & 0 & 0.89 & 1 \\ \hline
\tabincell{l}{after a languid intercontinental auction that stretched for more than a week, the american rights \\ to \textbf{jonathan\_littell}'s novel les bienveillantes, which became a publishing sensation in \textbf{france},\\ have been sold to harpercollins, the publisher confirmed yesterday.} & 0 & 0.11  & 0 \\ \hline
\end{tabular}
\caption{Instance selection examples by different models. For CNN+RL and CNN+Max, 1 or 0 means the sentence is selected or not. For CNN+ATT, the value is the attention weight.}
\label{tb:case}
\end{table*}

\vpara{Relation classification on selected data.} 
To measure the quality of the selected data by our instance selector, we performed relation classification experiments on the selected data. 
We first used our instance selector to select the high-quality sentences from the original data. Then, we trained two state-of-the-art models, CNN and CNN+ATT with two settings. One setting is to train them on the original data, named as CNN(Original) and CNN+ATT(Original). The other setting is to train them on the selected data, which are named as CNN(Selected) and CNN+ATT(Seleted). We compared the performance of CNN(Original) (CNN+ATT(Original)) with CNN(Selected) (CNN+ATT(Selected)) on the relation classification task. The results are compared under the held-out evaluation configuration~\cite{mintz2009distant} which provides an approximate measure of relation classification without expensive human annotations. The held-out evaluation compares the predicted relational fact from the test data with the facts in Freebase, but it does not consider the mapping between a relational fact and a sentence.

As shown in Figure~\ref{fig:cnn+rl} and Figure~\ref{fig:cnn+att+rl}, the models trained on the selected data achieve much better performance than the counterparts trained on the original dataset. The results also indicate our instance selector has the ability of filtering out noisy sentences and distilling high-quality sentences, resulting better classification performance.
 
\vpara{Accuracy of instance selection decision.} 
%The automatic evaluation show the effectiveness of the instance selector by conducted an relation classification experiment on the selected dataset.
To assess how accurate the decision is by the instance selector, we manually checked each sentence selected and rejected by the instance selector in a sampled dataset.
For each sentence, the instance selector makes a correct decision if the sentence's label is correct and our instance selector selects it as a training instance, or, if its label is wrong and our instance selector rejects it.
Otherwise, we judged that the instance selector makes a wrong decision. %% 

Specifically, we sampled $300$ sentences from the training data.
Our instance selector chooses $64$ sentences as the training instances, among which $45$ sentences are correctly selected. 
The selector also rejects $236$ instances, and $177$ of them are noisy instances (not describing the relation). 
To summarize, the accuracy of our instance selector is $(45+177)/300 = 74\%$, which demonstrates the effectiveness of our instance selector.

\vpara{Different instance selection strategies.}
To show the necessity for adopting the reinforcement learning framework for instance selection, we compared two instance selection strategies. Specifically, we performed relation classification on the selected data respectively with reinforcement learning (RL) selection and with greedy selection. The greedy selection selects the top $N$ sentences with the largest likelihood which is estimated by a pre-trained CNN.
%selection strategies.
%\begin{itemize}
%\item \textbf{Random Selection:} it randomly selects $N$ instances from the training data.
 %%
%\item \textbf{Greedy Selection:} it selects the top $N$ instances ranked by the likelihood estimated by a pre-trained CNN.
%\end{itemize}

%
\begin{figure}[t]
 \centering
   \includegraphics[width=0.33\textwidth]{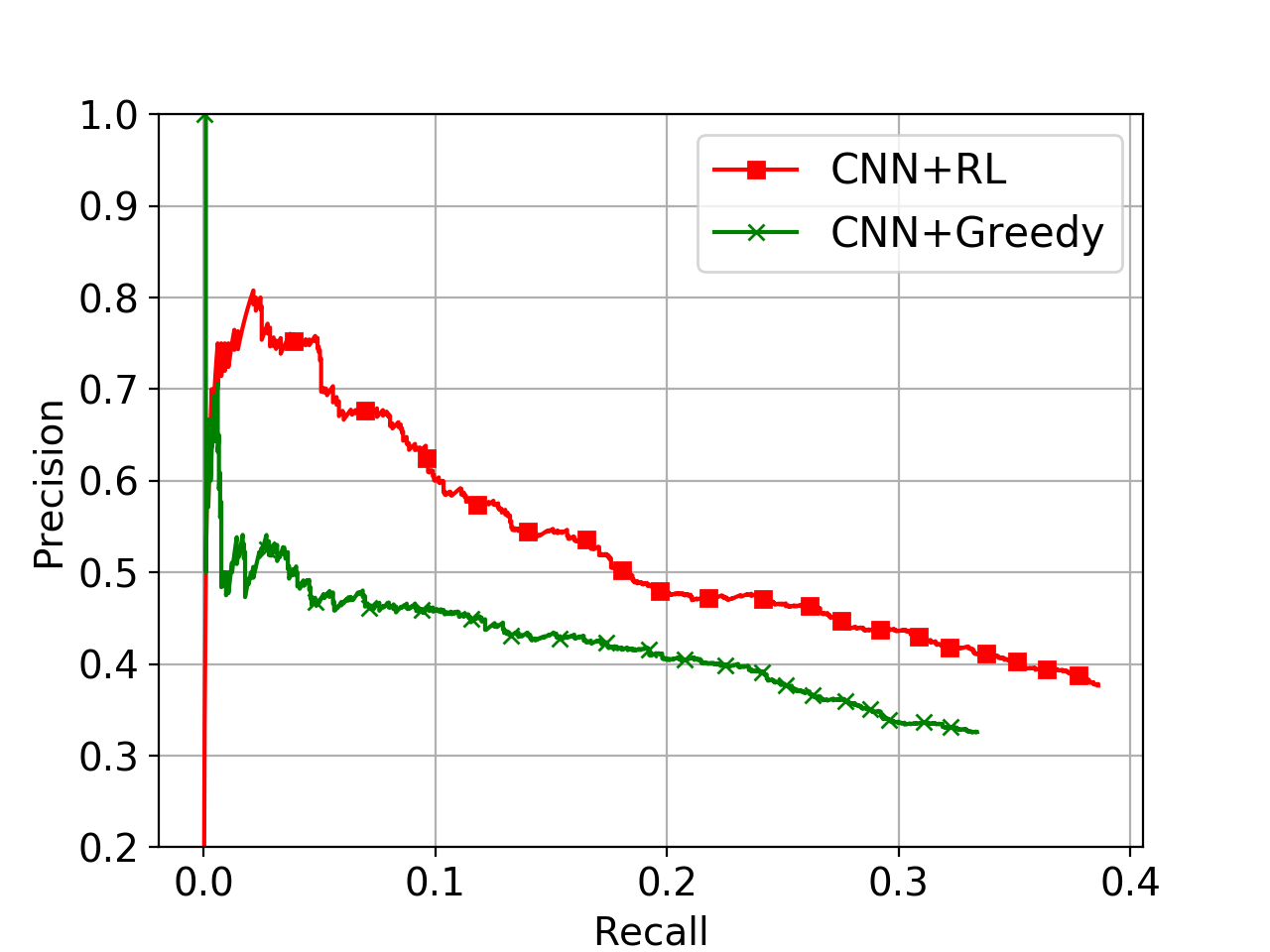}
   %\vspace{-0.1in}
   \caption{Comparison of instance selection with reinforcement learning against greedy selection. }
   \label{fig:greedy}
 %\vspace{-0.1in}	
\end{figure}
During the experiments, we kept the relation classifier untouched while replacing the RL selection by greedy selection. 
The number of selected instances $N$ is the same as the RL strategy. %% 
As shown in Figure~\ref{fig:greedy}, the performance of our instance selector is much better than the greedy strategy on the held-out evaluation. The results show that our RL strategy is reasonable and effective.

\vpara{Noisy bag filtering.}
As previous methods cannot filter the bags with all noisy sentences, we validated the ability of our model to filter bags with all noisy sentences. 
We randomly selected $100$ deleted sentence bags and find that $86\%$ of the bags consist of all noisy sentences.
This indicates that our instance selector can exclude the noisy sentences effectively.
%%
%Note that previous methods are unable to handle bags with all incorrect sentences.

\subsection{Case Study}
Table ~\ref{tb:case} shows two bag examples for instance selection. 
The first bag has three correct sentences. The second bag has two noisy sentences.
It is clearly show that our model can do better instance selection than both instance-weighting with CNN+ATT and maximum likelihood selection with CNN+Max.
The second example indicates that our model is able to filter bags with all noisy sentences while other methods fail to do so.

\section{Conclusion and Future Work}
In this paper, we propose a novel model for sentence-level relation classification from noisy data using a reinforcement learning framework.
%which are generated by distant supervised data.
The model consists  of  an  instance  selector and a relation classifier. The instance selector chooses high-quality data for the relation classifier. The relation classifier predicts relation at the sentence level and provides rewards to the selector as a weak signal to supervise the instance selection process.
Extensive experiments demonstrate that  our  model  can filter out the noisy sentences and  perform sentence-level relation classification better than state-of-the-art baselines from noisy data.
%Although the held-out evaluation already demonstrates the advantages of our proposal against the baselines,
%we are unable to annotate more data for manual evaluation at the time of submission,
%due to the heavy labor of manual annotation. In the future, we will annotate more data for manual evaluation.

Further, our solution for instance selection can be generalized to other tasks that employ noisy data or distant supervision. For instance, a possible attempt might be to perform sentiment classification on noisy data~\cite{Go2009Twitter}. 
We leave this as our future work. 

\section{Acknowledgement}
This work was partly supported by the National Science Foundation of China under grant No.61272227/61332007.

\bibliography{aaai}
\bibliographystyle{aaai}

\end{document}